# Coherent WDM transmission using quantum-dash mode-locked laser diodes as multi-wavelength source and local oscillator

Juned N. Kemal,[1] Pablo Marin-Palomo,[1] Vivek Panapakkam,[2] Philipp Trocha,[1] Stefan Wolf,[1] Kamel Merghem,[2] Francois Lelarge,[3] Abderrahim Ramdane,[2] Sebastian Randel,[1] Wolfgang Freude,[1] And Christian Koos [1,4,*]

[1]*Institute of Photonics and Quantum Electronics (IPQ), Karlsruhe Institute of Technology (KIT), Germany*
[2]*Centre de Nanosciences et de Nanotechnologies (CNRS), Univ. Paris-Sud, C2N - Université Paris-Saclay, France*
[3]*Now with Almae Technologies, Marcoussis, France*
[4]*Institute of Microstructure Technology (IMT), Karlsruhe Institute of Technology (KIT), Germany*
*\*Christian.koos@kit.edu*

**Quantum-dash (QD) mode-locked laser diodes (MLLD) lend themselves as chip-scale frequency comb generators for highly scalable wavelength-division multiplexing (WDM) links in future data-center, campus-area, or metropolitan networks. Driven by a simple DC current, the devices generate flat broadband frequency combs, containing tens of equidistant optical tones with line spacings of tens of GHz. Here we show that QD-MLLDs can not only be used as multi-wavelength light sources at a WDM transmitter, but also as multi-wavelength local oscillators (LO) for parallel coherent reception. In our experiments, we demonstrate transmission of an aggregate data rate of 4.1 Tbit/s (23×45 GBd PDM-QPSK) over 75 km standard single-mode fiber (SSMF). To the best of our knowledge, this represents the first demonstration of a coherent WDM link that relies on QD-MLLD both at the transmitter and the receiver.**

## 1. Introduction

Driven by the explosive growth of data traffic on all levels of optical communication networks [1], wavelength-division multiplexing (WDM) has become a necessity not only for long-haul transmission, but also for shorter links that connect, e.g., data centers across metropolitan or campus-area networks. These links crucially rely on compact WDM transceivers that offer multi-terabit/s aggregate data rates and that can be deployed in large quantities at low cost. In this context, optical frequency combs have emerged as a promising approach towards compact and efficient light sources that provide a multitude of tones for parallel WDM transmission [2-9]. Unlike the carriers derived from a bank of individual lasers, the tones of a comb are intrinsically equidistant in frequency and are defined by just two parameters – the center frequency and the free spectral range (FSR).

Among the different frequency comb sources, quantum-dash mode-locked laser diodes (QD-MLLD) are particularly interesting due to their technical simplicity: Driven by a DC current, these devices emit broadband frequency combs consisting of tens of optical carriers, spaced by free spectral ranges of, e.g., 25 GHz or 50 GHz, as determined by the cavity length. In previous experiments [6,9], the potential of QD-MLLD has been demonstrated by using the devices as multi-wavelength optical sources for parallel transmission on tens of WDM channels, achieving aggregate lines rates of more than 10 Tbit/s. In these experiments, symbol-wise phase-tracking [9] or self-injection locking [6] was used to overcome the limitations imposed by the rather high intrinsic linewidth of the QD-MLLD tones. However, while these demonstrations show the great potential of QD-MLLD as multi-wavelength light sources at the WDM transmitter, demodulation of the coherent signals still relied on a single-wavelength external-cavity laser (ECL), which acts as a local oscillator (LO) and needs to be tuned to the respective channel of interest. The scalability advantages of QD-MLLD for massively parallel WDM transmission have hence only been exploited at the transmitter, but not at the receiver side.

In this work, we demonstrate that QD-MLLD can also act as multi-wavelength local oscillators for intradyne reception of a multitude of optical channels [10]. In our experiments, we use a pair of QD-MLLDs with similar free spectral ranges (FSR) – one device to generate the optical carriers for the WDM transmitter, and the other device to provide the required LO tones for parallel coherent reception. We use 23 carriers separated by 50 GHz to transmit data at a symbol rate of 45 GBd using QPSK signaling, thereby obtaining data streams with an aggregate line rate of 4.14 Tbit/s and an aggregate net data rate of 3.87 Tbit/s which are transmitted over 75 km of standard single mode fiber (SSMF). To the best of our knowledge, this is the first demonstration of a QD-MLLD acting as a multi-wavelength LO in coherent transmission. When combined with highly integrated coherent transceiver circuits [11-14] and advanced digital signal processing (DSP) schemes implemented, e.g., in 7 nm CMOS technology [15,16], our approach could open a path towards coherent WDM transceivers with unprecedented compactness and scalability.

## 2. Optical frequency combs for WDM transmission and reception

The basic concept of comb-based WDM transmission and reception is shown in Fig. 1. The tones from the transmitter comb (Tx comb) are demultiplexed (DEMUX), and data is encoded onto the individual carriers by in-phase/quadrature modulators (IQ mod.). The optical channels are combined using a frequency multiplexer (MUX), amplified if needed, and transmitted through the transport fiber. At the WDM receiver, the of the MLLD via p-doped and n-doped InP layers, respectively. Lateral confinement of carriers is obtained by proton implantation [17]. Multiple longitudinal modes can coexist in the laser cavity due to the inhomogeneously broadened gain of the active region originating from the shape distribution of the QDs. Mode locking due to mutual sideband injection [18] produces an optical frequency comb output centered at 1545 nm with a line spacing of 25 GHz.

In coherent reception, it is desired to maintain a low frequency offset between the carrier of each channel and the corresponding LO tone at the receiver. This requires a Tx comb and an LO comb with similar center frequencies and free spectral ranges. To this end, we used two QD-MLLDs that are taken from the same wafer and are carefully cleaved to have very similar cavity lengths, thus resulting in overlapping spectra for appropriately set operating currents and temperatures. To select the operating conditions of the two QD-MLLDs, we channels are demultiplexed and sent to an array of independent coherent receivers (Coh. Rx). The demultiplexed spectral lines of a second frequency comb (LO comb) serve as local oscillator tones for the various coherent receivers.

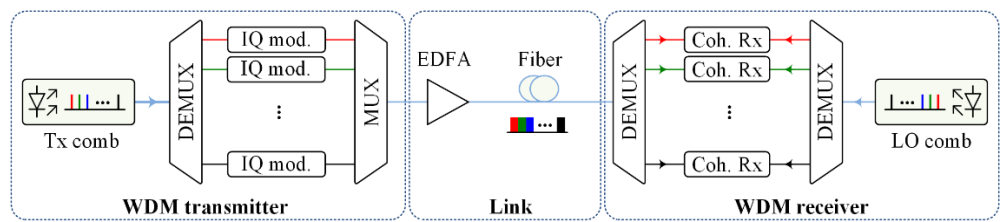

**Fig. 1.** Setup of a coherent WDM transmission link using optical frequency combs as light sources both at the transmitter and the receiver. WDM transmitter: Carriers from a transmitter comb (Tx comb) are demultiplexed (DEMUX) and modulated separately using in-phase/quadrature modulators (IQ mod.). The channels are then multiplexed (MUX) to form the WDM signal. Link: The WDM signal is amplified and transmitted. WDM receiver: The channels are demultiplexed and sent to an array of independent coherent receivers (Coh. Rx). The demultiplexed spectral lines of a local oscillator comb (LO comb) serve as LO tones for the various coherent receivers. EDFA: Erbium-doped fiber amplifier.

For our data transmission experiment, we use two QD-MLLDs – one as the Tx comb and the second as the LO comb. Figure 2(a) shows the basic structure of a QD-MLLD with an active region that consists of three stacked layers of InAs quantum dashes separated by InGaAsP barriers. Each QD layer is 2 nm thick and the barriers have a thickness of 40 nm. Two 80 nm-thick separate confinement heterostructure (SCH) layers of InGaAsP terminate the dash-barrier stack towards the top and bottom InP layers [17,18]. The optical mode is guided by a buried ridge waveguide of 1.0 μm width. Cleaved chip facets form a Fabry-Perot laser cavity with a length of 1.71 mm, leading to an FSR of 25 GHz. Top and bottom gold layers provide electrical contacts to the active region characterized the devices using the setup in Fig. 2(b). The emitted light is coupled to a lensed fiber, which is equipped with anti-reflection (AR) coating for 1550 nm and connected to an optical isolator to reduce optical back-reflection into the laser cavity. Different instruments, denoted by C, D, and E, are connected to the output of the isolator for measuring the data of Figs. 2 (c), 2(d), and 2(e), respectively.

Figure 2(c) shows the total output optical power of the Tx comb and the LO comb as a function of the injection current. Insertion losses of coupling interface to the lensed fiber (2 dB) and the isolator (1 dB) are taken into account. At a pump current of 300 mA, the output power radiated from the facet of the Tx comb exceeds 18 mW (12.5 dBm). The LO comb power peaks at 12 mW (10.8 dBm) for an injection current of 200 mA. We attribute the smaller output power of the LO comb to imperfect mounting and thermal coupling of the associated chip in our experiment.

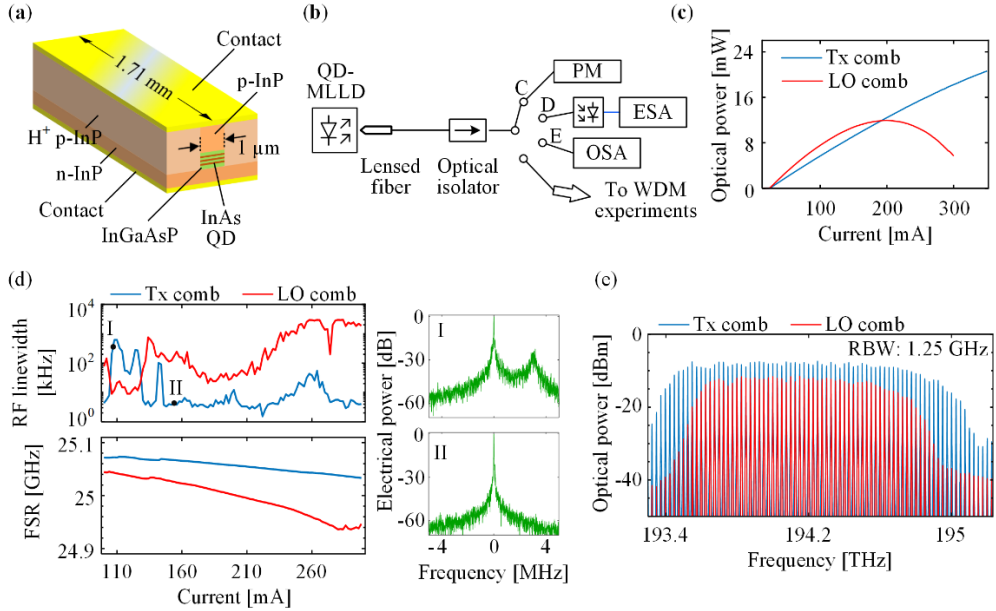

**Fig. 2.** Concept and characterization of quantum-dash mode-locked laser diodes (QD-MLLD). **(a)** Three-dimensional schematic of a QD-MLLD. The active region consists of three stacked layers of InAs quantum dashes, which are separated by InGaAsP barriers. The stack is terminated by a separate confinement heterostructure (SCH) layer to the top and the bottom. The buried ridge waveguide has a width of 1.0 µm. Cleaved chip facets form a Fabry-Perot laser cavity with a length of 1.71 mm, leading to an FSR of 25 GHz. Top and bottom gold layers provide electrical contacts to the active region of the MLLD via p-doped and n-doped InP layers. **(b)** Basic optical setup for using the QD-MLLD comb source. The device is driven by a constant injection current (not shown). The emitted light is collected by a lensed fiber, and an optical isolator is used to reduce back-reflections from the optical setup into the QD-MLLD. Setups C, D, and E are used for measuring the data of Subfigures (c), (d), and (e), respectively. PM: Power meter; ESA: Electrical spectrum analyzer; OSA: Optical spectrum analyzer. **(c)** Measured total comb power versus injection current of the Tx and the LO combs. Due to imperfect mounting of the LO QD-MLLD chip, the LO comb performs worse than the Tx comb. **(d)** Full-width half maximum (FWHM) of the radio-frequency (RF) beat note (upper plot) as a function of the drive current for the Tx comb (blue) and for the LO comb (red), and free spectral range (FSR, lower plot) as a function of the drive current for the Tx comb (blue), and for the LO comb (red). For some drive currents, we observe distinctively higher RF linewidths, see, e.g., Point I. In these cases, we often observe more than one beat note in the RF spectrum, Inset I, which we attribute to the coexistence of two sub-combs with slightly different FSR [20]. For such operating states, the depicted large RF linewidth values are to be understood as an indication of unstable operation by the MLLD. Inset II shows the measured RF beat note corresponding to stable single-mode operation of the QD-MLLD. In the insets, the spectral power is normalized to the dominant peak, and the frequency axis is defined with respect to the spectral position of this peak. **(e)** Optical power spectra of the Tx comb (blue) and the LO comb (red). The LO comb has a slightly lower bandwidth than the Tx comb. RBW: Resolution bandwidth of the OSA.

Figure 2(d), upper plot shows the measured FWHM of the RF beat note, commonly referred to as the RF linewidth, as a function of the injection current, both for the Tx comb and the LO comb. The RF beat note results from mixing neighboring comb lines in a photodiode and is measured by connecting the photodiode to an electrical spectrum analyzer (ESA), see Fig. 2(b), Setup D. A narrow RF linewidth indicates a reduced optical phase noise of the individual comb lines [19], which is advantageous for coherent optical communication. For certain drive currents, we observe comparatively large RF linewidths, see, e.g., Point I in Fig. 2(d). In these cases, we often observe more than one beat note in the RF spectrum, Inset I in Fig. 2(d), which we attribute to the coexistence of two sub-combs with slightly different FSR in the QD-MLLD cavity [20]. For such operating states, the large RF linewidth values refer to the dominant spectral peak and should only be taken as indicators for unstable operation by the QD-MLLD. Figure 2(d), Inset II shows the measured RF beat note corresponding to a stable single-mode operating state of the QD-MLLD. The lower plot in Fig. 2(d) shows the measured FSR of the Tx comb and the LO comb as a function of the injection current. During the above current sweeps, the sub-mounts of the Tx and LO MLLD chips are temperature-stabilized at 23.6 °C and 19.0 °C, respectively. These temperatures were chosen such that the tones of the two combs roughly coincide in center frequency. As can be seen in Fig. 2(d), the Tx comb has a small RF linewidth for a wide range of injection currents above 130 mA, whereas the LO comb

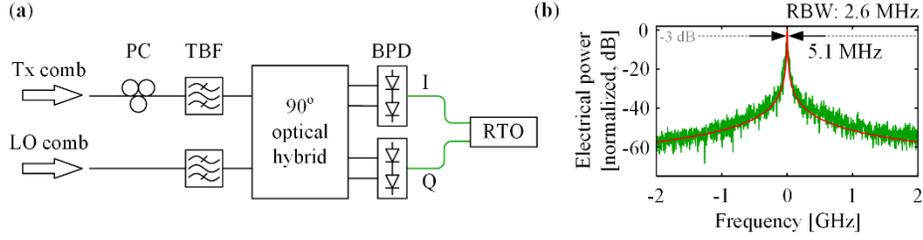

**Fig. 3**. Phase-noise characterization of QD-MLLD. **(a)** Experimental setup. Narrow-band tunable band-pass filters (TBF) are used to select distinct lines with comparable frequencies from the Tx comb and the LO comb. A polarization controller (PC) is used to match the polarizations of the selected tones at the input of a 90° optical hybrid. Balanced photodetectors (BPD) deliver the electrical in-phase (I) and quadrature (Q) signals, which are sampled with a real-time oscilloscope (RTO) for offline processing. **(b)** Power spectrum of the beat note between the two tones selected from the Tx comb and the LO comb as a function of the offset from the beat note frequency (green trace). The duration of the corresponding time-domain signals is 0.4 µs. We use the IQ data to independently extract a short-term Lorentzian linewidth of 5.1 MHz. The associated line shape (red trace) coincides well with the measured spectrum.

exhibits a small RF linewidth only for injection currents between 90 mA and 130 mA. This difference in behavior is again attributed to the imperfect mounting of the LO MLLD chip. For the subsequent WDM experiments with the two combs, we used injection currents of 278 mA and 125 mA for the Tx comb and the LO comb, respectively, giving nearly the same FSR (25 GHz) for both combs while maintaining low RF linewidths of less than 20 kHz. The optical spectra of the combs operated at the selected injection currents are shown in Fig. 2(e). The LO comb has a lower 3 dB bandwidth than the Tx comb, thereby limiting the total number of WDM channels that can be received.

The modulation formats and the symbol rates employed in a coherent communication link depend on the combined short-term linewidth of the transmitter carrier and the receiver LO [21]. Figure 3(a) shows the experimental setup used for estimating the linewidth of the beat note generated by mixing carrier and LO tones. Tunable band-pass filters (TBF) select single comb tones with comparable optical frequencies from the Tx comb and the LO comb. The selected tones are coupled to a 90° optical hybrid with a polarization controller in the Tx comb path to align the polarizations of the tones. The in-phase and quadrature outputs of the optical hybrid are detected with balanced photodetectors (BPD), and the electrical signals are sampled with two-channels of a real-time oscilloscope (RTO). Figure 3(b), green trace, shows the power spectrum of the beat note between a carrier tone of the Tx comb and the corresponding tone of the LO comb. The plot is obtained by a Fourier transform of the sampled time-domain data, which was recorded over a duration of 0.4 µs. The horizontal axis refers to the frequency offset from the center of the beat note.

For further analysis of the phase-noise characteristics, we extract the phases $\phi(t)$ of the recorded I/Q signal and use it for determining the combined short-term (Lorentzian) linewidth of the tones from the Tx and LO combs. To this end, the differences of the phases $\Delta\phi_\tau(t) = \phi(t+\tau) - \phi(t) = 2\pi f_i(t)\tau$ are calculated for various delay times $\tau > 0$ and the variance $\sigma^2_{\Delta\phi}(\tau)$ is computed. From this, the short-term Lorentzian linewidth $\Delta f$ is extracted from the slope of the variance $\sigma^2_{\Delta\phi}(\tau)$ at small time delays [22],

$$\Delta f = \lim_{\tau \to 0} \frac{\sigma^2_{\Delta\phi}(\tau)}{2\pi\tau} \ . \qquad (1)$$

This technique leads to a 5.1 MHz Lorentzian linewidth for the beat note. The same result is obtained by analyzing the FM noise spectrum and extracting the spectrally constant white-noise component [23]. The corresponding Lorentzian lineshape is plotted in Fig. 3(b), red trace, and shows good agreement with the directly measured lineshape, green trace. This indicates that, for the chosen observation time of 0.4 µs, the lineshape is dominated by high-frequency phase noise rather than by low-frequency drift. Since the estimated linewidth of the QD-MLLD tones is relatively high compared to that obtained from benchtop-type continuous-wave (CW) ECL, we use high symbol rates and employ symbol-wise phase tracking as in [9] for the transmission experiments discussed in Section 3.

## 3. Coherent optical communications using QD-MLLDs

### 3.1 Experimental setup for WDM transmission

For an experimental demonstration of coherent WDM transmission with QD-MLLDs as light sources both at the transmitter (Tx) and receiver (Rx), we use the setup depicted in Fig. 4(a). The total Tx comb output power of 9 dBm is boosted by an erbium-doped fiber amplifier (EDFA-1). The spacing between the amplified comb lines

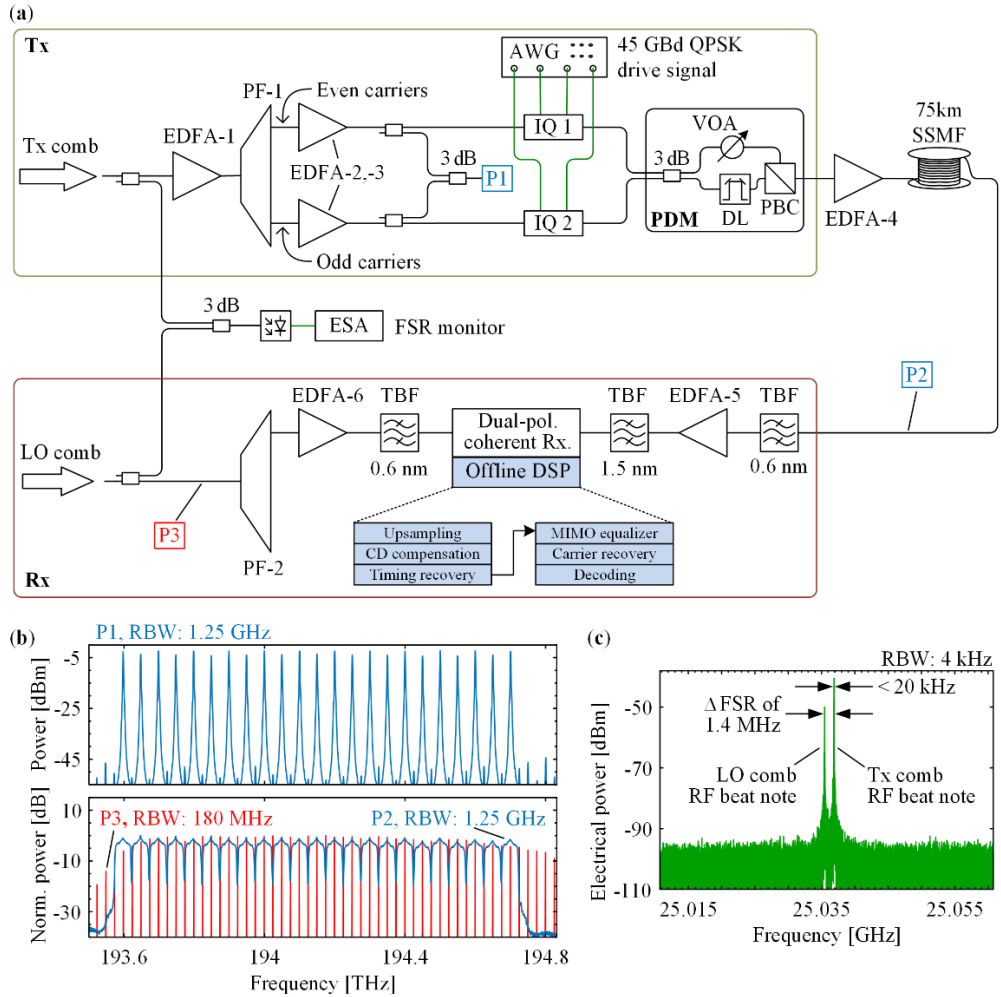

**Fig. 4.** Coherent WDM transmission experiment using QD-MLLD comb generators both as a multi-wavelength light source at the transmitter (Tx) and as a multi-wavelength LO at the receiver (Rx). **(a)** Experimental setup: The Tx comb is de-interleaved into even and odd carriers using a programmable filter (PF-1). The two sets of carriers are then modulated with independent data streams to emulate a real WDM signal with independent data on neighboring channels. Polarization-division multiplexing (PDM) is emulated by splitting the single-polarization WDM signal into two paths, delaying one of them, and recombining the two decorrelated signals on orthogonal polarizations of the transmission fiber. At the receiver, we subsequently select individual lines of an LO comb and use them for coherent detection of the corresponding WDM channel. The WDM channels are spaced by approximately 50 GHz and carry QPSK signals at symbol rates of 45 GBd, limited by the bandwidth of the arbitrary-waveform generator (AWG) used to generate the modulator drive signals at the Tx. P1, P2, and P3 denote points in the experimental setup at which the optical spectra in Subfigure (b) have been taken. EDFA: Erbium-doped fiber amplifier, IQ: In-phase/quadrature modulator; DL: Delay line; VOA: Variable optical attenuator; PBC: Polarization beam combiner; TBF: Tunable band-pass filter; SSMF: Standard single-mode fiber; ESA: Electrical spectrum analyzer; CD: Chromatic dispersion; MIMO: Multiple-input and multiple-output. **(b)** Comb spectra of both even and odd carriers before modulation (upper plot, P1), spectrum of 23 modulated carriers (lower plot, P2), and spectrum of the LO comb (lower plot, P3). **(c)** RF beat note spectrum resulting from the self-beating of nearest-neighbored comb lines of each comb when detected directly with a photodetector. The spectrum was obtained by impinging the output of the two QD-MLLDs on a photodetector simultaneously as depicted in Subfigure (a).

is increased to 50 GHz by selecting every second line with a programmable filter (PF-1, Finisar) such that a symbol rate of 45 GBd can be used, see spectrum in Fig. 4(b) which was taken at point P1 of the experimental setup. The programmable filter PF-1 is additionally used to flatten the comb spectrum. We emulate WDM traffic by encoding independent data streams on neighboring carriers. This is done by de-interleaving 23 spectral lines from the Tx comb into even and odd carriers using PF-1. The even and odd carriers are amplified by EDFA-2 and

EDFA-3, respectively. Both carriers are then independently modulated with pseudo-random bit sequences (PRBS) of length $2^{11} - 1$. This bit sequence is mapped to QPSK symbols with a rate of 45 GBd and a raised-cosine spectrum having a 10 % roll-off. In our experiments, the symbol rate was limited by the bandwidth of the arbitrary-waveform generator (AWG) used to generate the drive signals for the IQ modulators at the transmitter. The modulated even and odd carriers are then recombined by a polarization-maintaining (PM) 3 dB coupler to form a WDM signal with uncorrelated data streams on neighboring channels. To emulate polarization-division multiplexing (PDM), the WDM signal is split into two paths. In one of the paths, the signal is delayed by about 5.34 ns (240 symbols) with an optical delay line (DL) for decorrelation. A variable optical attenuator (VOA) in the other path equalizes the signal powers in the two paths. The signals are then recombined in orthogonal polarization states using a polarization beam combiner (PBC). The dual-polarization WDM signal is amplified and transmitted over a 75 km SSMF link, see spectrum in Fig. 4(b), which was taken at point P2 of the experimental setup. At the receiver, the WDM channel of interest is selected using a 0.6 nm tunable band-pass filter (TBF) and amplified by EDFA-5. A second 1.5 nm-wide TBF is used to suppress out-of-band amplified spontaneous emission (ASE) noise of EDFA-5. The selected channel is detected by a dual-polarization coherent receiver (Keysight N4391A). The corresponding reference tone for coherent detection is selected from the LO comb using a second programmable filter (PF-2), amplified by EDFA-6, and filtered by another 0.6 nm TBF for ASE noise suppression. The output of the coherent receiver is digitized using four oscilloscope channels with a sampling rate of 80 GSa/s each (Keysight DSOX93204A) and evaluated offline using digital signal processing (DSP). In the DSP, we first up-sample the received signal to two samples per symbol. This is followed by blind chromatic-dispersion compensation based on the Godard clock-tone algorithm [24] after which timing recovery is carried out [25,26]. An adaptive multiple-input and multiple-output (MIMO) equalizer using the constant-modulus algorithm (CMA) then demultiplexes the two orthogonal polarizations of the signal and compensates linear transmission impairments [27]. We use periodogram-based techniques to compensate for frequency offsets between the signal carrier and the LO tone. Subsequently, carrier phase noise compensation is carried out based on a blind phase search (BPS) algorithm [21]. This technique allows for symbol-wise phase tracking and has been shown to effectively cope with the phase noise of QD-MLLD carriers in previous experiments [9]. Figure 4(b) depicts the superposition of the LO comb tones with the WDM signal. As discussed in Section 2, the FSR of the combs are nearly identical and the modulated channels are hence in good frequency alignment with the corresponding LO tones. Figure 4(c) shows the RF beat notes of the Tx comb and the LO comb captured simultaneously using an electrical spectrum analyzer (ESA, Keysight PXA N9030A), see "FSR monitor" in Fig. 4(a). The two combs have narrow RF linewidths (< 20 kHz) and FSRs with a very small difference ($\Delta$ FSR = 1.4 MHz).

### 3.2 WDM transmission over 75 km at a net data rate of 3.8 Tbit/s

The results of our transmission experiment are shown in Fig. 5. Throughout our experiments, the BER was too low to be measured within a limited record length of $1\times0^6$ bits. We therefore rely on the error-vector magnitude (EVM, normalized to the maximum constellation amplitude) to quantify signal quality and estimate the BER assuming that, after phase-noise compensation, the QPSK signals are impaired by additive white Gaussian noise only [28]. The measured EVM for each WDM channel is depicted in the upper plot of Fig. 5(a), and the corresponding BER estimates are shown in the lower plot. All channels fall well below the BER threshold of $4.5\times10^{-3}$ for hard-decision forward-error correction (HD-FEC) with 7% overhead [29]. This is confirmed by the fact that none of our signal recordings of $1\times10^6$ bits each showed any error. This leads to an aggregate line rate of 4.14 Tbit/s and a net data rate of 3.87 Tbit/s, transmitted over 75 km of SSMF. Example constellation diagrams obtained for the channel at 194.15 THz, both for back-to-back and for fiber transmission are shown in Fig. 5(b). The PRBS length used in these experiments was $2^{11}- 1$.

### 3.3 Comparison between QD-MLLD LO and ECL LO

In this section, we investigate whether or not there is a penalty in received signal quality when using a QD-MLLD comb line as LO rather than an ECL LO. The experimental setup is depicted in Fig. 6(a) and relies on an ECL as light source at the Tx. The ECL carrier, having a linewidth of less than 100 kHz and a power of 6 dBm, is amplified by EDFA-1 prior to modulation with a QPSK signal at a symbol rate of 45 GBd. In this experiment, we used PRBS of length $2^{11} - 1$ and $2^{15} - 1$, both of which lead to identical results. In the following, we show only the results for the longer sequence. PDM is emulated as discussed in Section 3.1. The PDM signals are amplified by EDFA-2, filtered by a TBF to remove ASE noise, and sent to the signal input of a dual-polarization coherent receiver. The modulated signal is received by using either a QD-MLLD comb line or a tone obtained from a second ECL as LO. The LO power is boosted by EDFA-3 and filtered by PF, which is mainly needed for the comb input. We vary the optical carrier-to-noise power ratio (OCNR) of the LO tone using a variable optical attenuator (VOA) in front of another EDFA (EDFA-4). The OCNR for a reference noise bandwidth of 12.5 GHz (0.1 nm) is measured at the monitor output of the dual-polarization coherent receiver (Keysight N4931A). Offline DSP as described in Section 3.1 is employed to analyze the received data. Figure 6(b) shows the measured BER with $1\times10^6$ analyzed bits versus the OCNR of the LO. Note that, due to the limited number of $10^6$ analyzed bits, the BER values below $10^{-5}$ might be subject to statistical inaccuracies. Different LO tones from the QD-MLLD are tested by using the corresponding Tx carrier frequency. As can be seen from Fig. 6(b), LO tones from a QD-MLLD (solid lines) perform similarly to an ECL LO (dotted lines). However, when using QD-MLLD LO tones at low OCNR, cycle slips were observed in the recovered symbols due to incorrect estimation of the carrier phase by a multiple of π/2. The impact of this cycle slips can be reduced by using differential coding along with phase-slip tolerant FEC schemes [30] or by adding pilot symbols for accurate phase referencing [31].

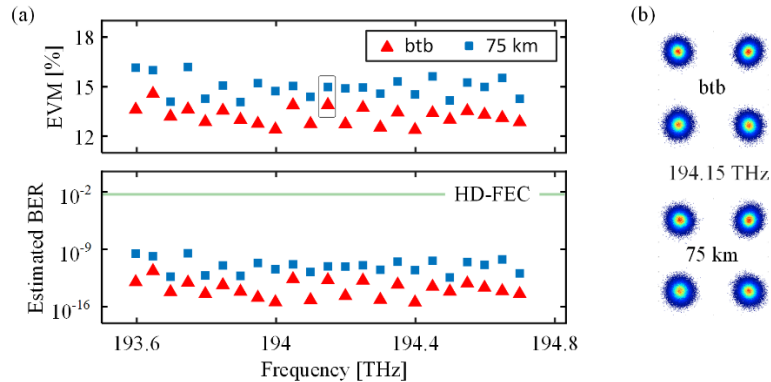

**Fig. 5.** Coherent data transmission performance obtained by using a QD-MLLD comb generator both as the Tx light source and as the LO. **(a)** Upper plot: Measured EVM as a function of the carrier frequency of the transmitted channels for both back-to-back (btb, ▲) and 75 km SSMF transmission (■). Lower plot: BER estimated from EVM assuming that the QPSK signals are impaired by additive white Gaussian noise only [28]. **(b)** Example constellation diagrams for the signal carried by the comb line at 194.15 THz, both for back-to-back (btb) and for 75 km fiber transmission.

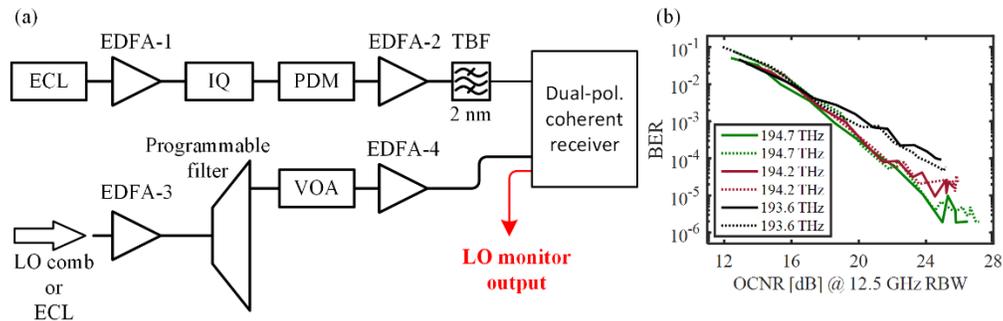

**Fig. 6.** Comparison of transmission performance obtained by using a QD-MLLD and a conventional external-cavity lasers (ECL) as LO tone generators. **(a)** Experimental setup. At the Tx, we use an ECL having a linewidth below 100 kHz as light source. The ECL carrier is boosted by an EDFA (EDFA-1) and modulated with a QPSK signal at a symbol rate of 45 GBd. The length of the underlying pseudo-random bit sequence amounts to $2^{15}-1$. Polarization-division multiplexing (PDM) is emulated by the same method discussed in Section 3.1. The PDM signals are amplified by an EDFA (EDFA-2), filtered by a tunable bandpass filter (TBF), and sent to the signal input of a dual-polarization coherent receiver. The signal is detected by using either a QD-MLLD comb tone as an LO, or by using a second ECL. The LO is amplified by EDFA-3 and filtered by a programmable filter. A variable optical attenuator (VOA) in front of EDFA-4 is used to vary the optical carrier-to-noise power ratio (OCNR) of the LO tone. **(b)** Plot of BER vs. OCNR of the LO tone. Results obtained from the ECL are depicted as dotted lines, whereas the solid lines refer to the QD-MLLD LO tones. The OCNR is indicated with respect to a reference noise bandwidth of 12.5 GHz (0.1 nm). We find that LO tones from a QD-MLLD (solid lines) perform similarly to an ECL LO (dotted lines).

## 4. Summary

We demonstrate that QD-MLLD can be used both as multi-wavelength light sources at the Tx and as multi-wavelength LO at the Rx of massively parallel WDM systems. In out experiments, we use 23 carriers to transmit a line rate (net data rate) of 4.140 Tbit/s (3.869 Tbit/s) over a 75 km standard single mode fiber (SSMF) link. To the best of our knowledge, this is the first time that an optical frequency comb generated by a QD-MLLD has been used as a multi-wavelength LO in a coherent transmission system. Due to their compactness and simple operation requiring a DC drive current only, QD-MLLD represent a particularly interesting comb generator concept for efficient and highly scalable WDM transceivers.

**Funding and Acknowledgments**

This work was supported by the European Research Council (ERC Starting Grant 'EnTeraPIC', # 280145, ERC Consolidator Grant 'TeraSHAPE', # 773248), the EU project BIG PIPES (# 619591), by the Alfried Krupp von Bohlen und Halbach Foundation, by the Helmholtz International Research School for Teratronics (HIRST), and by the Karlsruhe School of Optics & Photonics (KSOP). We gratefully acknowledge support from the Erasmus Mundus Joint Doctorate program EUROPHOTONICS (Grant No. 159224-1-2009-1-FR-ERA MUNDUS-EMJD) and from the Deutsche Forschungsgemeinschaft (DFG) through CRC 1173.